\documentclass[english,aps,prb,floatfix,twocolumn,showpacs]{revtex4}
\usepackage{amsmath}
\usepackage{babel}
\usepackage{graphicx}
\usepackage{amssymb}
\usepackage{multirow}

\begin{document}

\title{Understanding ferromagnetism in Cr-based 3d-5d double perovskites}
\author{Prabuddha Sanyal} 
\affiliation{Center for Basic Sciences, University of Mumbai}

\date{\today}

\begin{abstract}
Ferromagnetism in Cr-based double perovskites is analyzed using effective model as well as simulation approach.
 Starting from a microscopic
model proposed recently for this class of double perovskites, 
 an effective spin-only model is derived in the limit of large
exchange coupling at the B-site. Analytic expressions for the resultant exchange coupling is derived in certain limiting cases. 
 The behaviour of this exchange is used to provide a plausible explanation for the enhanced ferromagnetic tendency as also the enigmatic increase in Curie temperature observed in Cr-based DP-s,
in the series Sr$_{2}$CrWO$_{6}$,Sr$_{2}$CrReO$_{6}$ to Sr$_{2}$CrOsO$_{6}$. The superexchange between neighbouring B and B' sites is found to play
a crucial role both in stabilizing ferromagnetism, especially in the latter two compounds, as well as increasing the T$_{c}$.

\end{abstract}
\maketitle
\noindent
\section{Introduction}

In recent times, ferromagnetic transition metal compounds with high $T_{c}$ have come into the focus for their
importance in spintronics and other technological applications.  While doped rare earth manganites have been
 studied for decades for their CMR property, no less well known are the double perovskites,
 with a general formula A$_{2}$BB$'$O$_{6}$, A=alkaline/rare-earth metals and B/B$'$=transition metals. The most well known member, Sr$_{2}$FeMoO$_{6}$ has a $T_{c}$ of 410K, higher than most
manganites~\cite{SFMO,DDreview}. Moreover, the  half-metallic nature~\cite{tomioka,TSD}, coupled with the substantial tunnelling magnetoresistance obtained in these compounds at low temperatures, especially
when powdered, makes it valuable for spintronic applications~\cite{Mag-Res1,Mag-Res2,Mag-Res3}. The high $T_{c}$ implies a strong polarization even
at room temperatures, enhancing its industrial importance. In an effort to boost the $T_{c}$ even further, researchers have electron doped this compound~\cite{JAP,ES,Navarro,photoemission}. Although the $T_{c}$ does increase in this process upto some point~\cite{Navarro,p21n}, but it has been shown in a recent work that upon overdoping,
the $T_{c}$ actually decreases, and the ferromagnetism becomes unstable~\cite{mePinaki,LSFMO}! In fact, it gets
replaced by antiferromagnetic phases.  
This conclusion had been derived both within model Hamiltonian approach~\cite{mePinaki} as well as within abinitio approach. 
 Within the model approach, 
it was observed that the
effective exchange interaction between Fe $S=5/2$ core spins changes sign as the filling increases, signalling a crossover from ferro to antiferro. Within the abinitio approach, considering the Sr$_{2-x}$La$_{x}$FeMoO$_{6}$ series~\cite{LSFMO}, it was observed that  
  the ferromagnetic state became progressively unstable as the number of valence electrons was increased through increased doping of La, the balance tilting at an electron count 
of about $2.4$. While the La-overdoped regime of this compound has not been investigated in detail as yet, recent experimental data support our claims regarding the disappearance of
ferromagnetism on electron doping~\cite{Sugatapriv}. Hence simple electron doping by
A-site cataion substitution is not a very promising method for increasing $T_{c}$ beyond
a point. There is however, another well known technique to change the electron filling, namely
 to substitute instead the $B^{\prime}$ site ion. In this case, it is well known that in
 Cr-based double perovskites: Sr$_{2}$CrB$^{\prime}$O$_{6}$, as one goes across the period,
 substituting $B^{\prime}$ in turn as W, Re and Cr, the $T_{c}$ increases progressively,
 reaching a high of over 700K for the Os compound. Obviously, the ferromagnetism is not
 becoming unstable in this series of compounds even though the filling is increasing from
 $1$ to $2$ to $3$ as one goes from W to Re to Os respectively. On the contrary, it is becoming more stabilized, as signified by the increasing $T_{c}$. 
In this
paper, we probe the role of a novel superexchange mechanism, in addition to the kinetic energy driven mechanism already prevalent, to account for this anomalous stabilization of ferromagnetic behaviour. While abinitio and variational approaches had earlier~\cite{3d5d} provided a pointer to this superexchange mechanism,
 in this
communication, we conclusively demonstrate its importance using effective exchange calculations and direct numerical simulations. The organization of the paper is as follows. In the next section, we shall briefly summarize the Hamiltonian as well as main
results for the Sr$_{2-x}$La$_{x}$FeMoO$_{6}$ series (i.e., cataion site substitution), just to serve as a reminder, and for constrast with the Cr-based series (B-site substitution). Next,
we provide a brief account of the main results of abinitio studies upon these compounds, which have been reported in detail earlier. In the fourth section, we motivate a new modified Hamiltonian for this Cr-series. We proceed to derive a low-energy, spin-only model from this fermionic
Hamiltonian in the next section, and analyze its magnetic properties. 

\section{A-site Cataion doping: Brief summary}

A-site cataion substitution can be done upon the parent compound Sr$_{2}$FeMoO$_{6}$ using,
for example La which has a nominal valence state of $3+$, in place of Sr which has a valence
state of $2+$. This would correspond to electron doping of the system. Abinitio studies
on this series of compounds~\cite{LSFMO}, using NMTO~\cite{NMTO} downfolding have shown that the
relative positions of Fe and Mo $t_{2g}$ orbitals remain almost unchanged upon La doping,
only the Fermi energy shifts almost like a rigid band picture. Thereupon, total energy calculations using Vienna Abinitio Simulation Package (VASP)~\cite{VASP} showed that the magnetic ground state changed from ferro to antiferro as the electron doping increased. 
The spin splitting at the Fe and Mo site was also obtained by downfolding everything
except the $t_{2g}$ orbitals. It was observed that the spin-splitting at the Mo-site increased proportional
to the filling, so that the Stoner $I$ remained essentially constant. For example,
in Sr$_{2}$FeMoO$_{6}$ which has 1 electron per site the spin-splitting at the Mo site is about 0.13 $eV$, while in La$_{2}$FeMoO$_{6}$, with a filling of 3, the splitting is 0.37 $eV$.  This showed that
the moment at the Mo site is completely induced from the Cr spin, and not intrinsic. 
It was shown in that paper
that the Hamiltonian which captures this behaviour is given by:

$$H= \epsilon_{Fe}\sum_{i\in B}f_{i\sigma\alpha}^{\dagger}f_{i\sigma\alpha}+
\epsilon_{Mo}\sum_{i\in B'}m_{i\sigma\alpha}^{\dagger}m_{i\sigma\alpha} $$

$$-t_{FM}\sum_{<ij>\sigma,\alpha}f_{i\sigma,\alpha}^{\dagger}m_{j\sigma,\alpha} 
-t_{MM}\sum_{<ij>\sigma,\alpha}m_{i\sigma,\alpha}^{\dagger}m_{j\sigma,\alpha} $$ 
\begin{eqnarray}
-t_{FF}\sum_{<ij>\sigma,\alpha}f_{i\sigma,\alpha}^{\dagger}f_{j\sigma,\alpha} 
+ J_{1}\sum_{i\in A} {\bf S}_{i} \cdot
f_{i\alpha}^{\dagger}\vec{\sigma}_{\alpha\beta}f_{i\beta}
\label{fullhamSFMO}
\end{eqnarray}

The $f$'s refer to the Fe sites and the $m$'s to the Mo sites.
$t_{FM}$, $t_{MM}$, $t_{FF}$ represent the nearest neighbor Fe-Mo, second
nearest neighbor Mo-Mo and Fe-Fe hoppings respectively, the largest hopping being given
by $t_{FM}$.  $\sigma$ is the spin index and $\alpha$ is the orbital index that spans the t$_{2g}$ 
manifold. The difference between the ionic levels,
${\tilde \Delta} = \epsilon_{Fe} - \epsilon_{Mo}$, defines the charge transfer energy.
Since among the crystal-field split d levels of Fe and Mo, only the relevant $t_{2g}$ orbitals are retained,
giving rise to on-site and hopping matrices of dimension 3 $\times$ 3. 
The ${\bf S}_i$ are `classical' (large $S$)
 core spins at the B site, coupled
to the itinerant B electrons through a coupling $J_{1} \gg t_{FM}$. 
Variants of this two-sublattice Kondo lattice model has been considered by several authors~\cite{Millis,Avignon,Guinea1,Guinea2,Navarro}in the context of double perovskites.

The abinitio calculations showed that the charge transfer energy $\Delta$ remains almost
fixed upon La-doping. Consequently, the doping only corresponds to increasing the Fermi energy
keeping the levels fixed, thereby increasing the electron filling.  
   Thus, the ground state changes from ferromagnetic to
  antiferromagnetic at a filling of about 2.2-2.4 as shown in Ref~\cite{mePinaki}..

\section{B$^{\prime}$ site substitution: Main abinitio results}

  In the series of compounds Sr$_{2}$CrB$^{\prime}$O$_{6}$, the B$^{/}$ ion which is in nominal 
 5+ valence state corresponds to $5d^{1}$,$5d^{2}$,$5d^{3}$ configuration of W, Re and Os respectively. These materials have been reported to have $T_{c}$ as high as about 450K, 620K and 725K.
 with a progressive increase as one moves from Sr$_{2}$CrWO$_{6}$ with one valence electron to
 Sr$_{2}$CrOsO$_{6}$ with three valence electrons. Taking the number of valence electrons as
 sole consideration, the situation of Sr$_{2}$CrWO$_{6}$,Sr$_{2}$CrReO$_{6}$,Sr$_{2}$CrOsO$_{6}$ is comparable to Sr$_{2}$FeMoO$_{6}$, SrLaFeMoO$_{6}$ and La$_{2}$FeMoO$_{6}$. However, unlike  the Cr-B$^{/}$ ($B^{/}$=W,Re,Os) series, for the Sr$_{2-x}$La$_{x}$FeMoO$_{6}$, the ferromagnetic $T_{c}$ was found to decrease with increasing La concentration (i.e., increasing number of valence electrons) and  finally the antiferromagnetic phase taking over the ferromagnetic phase.
 The Cr-B$^{/}$ (B$^{/}$=W,Re,Os) series though bear two fundamental differences compared to 
Sr$_{2-x}$La$_{x}$FeMoO$_{6}$ series. Firstly, the B$^{/}$ ions in Cr-B$^{/}$ being 5d transition metals exhibit significant spin-orbit coupling which make these materials suitable for magneto-optic applications with large signal as has been discussed in Ref1. Secondly, three
 different chemical elements, namely W, Re and Os are involved in Cr-B$^{/}$ series while for 
 Sr$_{2-x}$La$_{x}$FeMoO$_{6}$ series the increased electron count is achieved without any
 changes in the B-B$^{/}$ sublattice. First principle calculations using generalized
gradient approximation (GGA)~\cite{GGA}, carried out using  
(VASP) as well as Linear Muffin Tin Orbital (LMTO) have been carried out for the
total energy calculations. Thereafter, a few-band, low-energy tight binding Hamiltonian
was derived using N-th order muffin-tin orbital (NMTO)~\cite{NMTO} formalism. Details have been
reported elsewhere~\cite{3d5d}; here we recapitulate the main results. 

It was observed that the moment on the $B^{\prime}$ site increased from $0.3\mu_{B}$ for 
tungsten (W) to $0.81\mu_{B}$ for Rhenium (Re) to $1.44 \mu_{B}$ for Osmium. The total
moment, on the other hand, goes from $2\mu_{B}$ in W to $1\mu_{B}$ in Re to $0\mu_{B}$ in Os.
It is observed that although the total moment decreases in steps commensurate with the
filling, and the Cr moment remains almost fixed, the moment at the $B^{\prime}$ site
increases drastically, pointing to the growing localization of electrons on this site.

To probe this issue in more detail, the Sr, Oxygen orbitals as well as the $e_{g}$ orbitals
of Cr and $B^{\prime}$ ions were downfolded using NMTO formalism, keeping only the 
Cr and $B^{\prime}$ $t_{2g}$ orbitals. It was observed that the spin splitting at the
$B^{\prime}$ site increased from 0.06 $eV$ at the W site to 0.31 at the Re site to 0.53 at the Os
site. This is obviously a much faster increase than the filling would dictate, indicating that
the Stoner $I$ is itself increasing. If we multiply the splitting for W, i.e., 0.06 by 2 and
3 respectively, then the extra amount must correspond to a different energy scale in the
 problem. Let us call this energy scale as $J_{2}$, while the traditional 
extremely large spin splitting
 at the B site, or Chromium site defines the other exchange energy scale, $J_{1}$. 

Another important point which emerges out of abinitio calculations is the progressive increase
in the charge transfer energy $\Delta$ as one goes across the series from W to Re to Os.
It increases from 0.51 $eV$ in W, to 0.9 in Re, to 1.35 in Os. This situation is to be
 contrasted with the nearly constant $\Delta$ in the Sr$_{2-x}$La$_{x}$FeMoO$_{6}$ series.
 Thus, it may be summarized that both $\Delta$ and $J_{2}$ are increasing in the Cr-series
 from W to Re to Os compounds, although $J_{2}$ is usually much less than $\Delta$.

\section{Hamiltonian}

From the above considerations, it is obvious that there exists a different exchange energy scale
to the problem, apart from the spin splitting at the Cr site $J_{1}$, which, being
related to the Hund coupling, for all practical
purposes can be considered infinite. The increasing localized character of the moment on the
$B^{\prime}$ site makes superexchange an obvious candidate for this new energy scale~\cite{GuyCohen}. 
Hence we have added a  superexchange term to the Hamiltonian corresponding to superexchange
between the $t_{2g}$ spin on the $B$ site, and the electron spin on the $B^{\prime}$ site.
Hence,the representative Hamiltonian is given by:

$$H= \epsilon_{Fe}\sum_{i\in B}f_{i\sigma\alpha}^{\dagger}f_{i\sigma\alpha}+
\epsilon_{Mo}\sum_{i\in B'}m_{i\sigma\alpha}^{\dagger}m_{i\sigma\alpha} $$
$$-t_{CB^{/}}\sum_{<ij>\sigma,\alpha}f_{i\sigma,\alpha}^{\dagger}m_{j\sigma,\alpha} 
-t_{B^{\prime}B^{/}}\sum_{<ij>\sigma,\alpha}m_{i\sigma,\alpha}^{\dagger}m_{j\sigma,\alpha} $$ 
$$-t_{CC}\sum_{<ij>\sigma,\alpha}f_{i\sigma,\alpha}^{\dagger}f_{j\sigma,\alpha} 
+ J_{1}\sum_{i\in A} {\bf S}_{i} \cdot
f_{i\alpha}^{\dagger}\vec{\tau}_{\alpha\beta}f_{i\beta}$$
\begin{eqnarray}
+J_{2}\sum_{i\in B} {\bf S}_{i} \cdot
m_{i\alpha}^{\dagger}\vec{\sigma}_{\alpha\beta}m_{i\beta}
\label{ham3d5d}
\end{eqnarray}

 This spin-fermion Hamiltonian is in general difficult to solve exactly, but some headway
 in the direction of understanding the low-energy magnetic behaviour can be made if
 on can obtain a spin-only model from this with an effective exchange. In principle this
 should be possible by tracing out the fermion degrees of freedom, but that is a herculean
 task considering the multitude of possible configurations. Instead, we consider the
 approximate but enlightening procedure of Self-Consistent Renormalization (SCR) devised
  by Kumar and Majumdar~\cite{PinakiSCR}. 

\section{Derivation of effective spin model}

In an earlier paper~\cite{mePinaki}, I had obtained an effective spin model from the two-sublattice Kondo lattice model(Eqn~\ref{fullhamSFMO}) appropriate for the Fe-series (Sr$_{2-x}$La$_{x}$FeMoO$_{6}$) using
the procedure of SCR. In the action corresponding to this Hamiltonian, all but the last
 term is translationally invariant, hence can be fourier transformed and written in terms of the bare Green's function. Thereafter, the Molybdenum degrees of freedom can be integrated out,
 and the Iron degrees of freedom traced over, after taking the limit of $J_{1}\rightarrow\infty$,
 to obtain an effective model containing only the Iron spins.
 One may imagine that the same procedure may be followed in this case. There is a fundamental
 difficulty in this case though, due to the fact that there is spin-disorder even on the
 Molybdenum site, making it impossible to diagonalize and integrate out the Molybdenum
 degrees of freedom. Hence, the entire procedure of thinning of degrees of freedom has to
 be done in real space. The action corresponding to the Hamiltonian in Eqn~\ref{ham3d5d}
is given by:

$$\mathcal{A}=\sum_{i\omega_{n}} \left[ (i\omega_{n}-\epsilon_{Fe})\sum_{i\sigma}f_{i,n,\sigma}^{\dagger}f_{i,n,\sigma}+(i\omega_{n}-\epsilon_{Mo})\sum_{i\sigma}m_{i,n,\sigma}^{\dagger}m_{i,n,\sigma}\right.$$
\begin{eqnarray}
+t_{FM}\sum_{<ij>}(f_{i,n,\sigma}^{\dagger}m_{j,n,\sigma}+h.c.) \nonumber\\
\left.+J_{1}\sum_{i\alpha\beta}\vec{S}_{i}\cdot f_{i,n,\sigma}^{\dagger}\vec{\tau}_{\alpha,\beta}f_{i,n,\beta}+J_{2}\sum_{i}\vec{S_{i}}\cdot m_{i,n,\sigma}^{\dagger}\vec{\sigma}_{\alpha\beta}m_{i,n,\beta}\right]
\label{action3d5d}
\end{eqnarray}
Now, since the spin orbit coupling at the $B^{\prime}$ site is large for the 5d elements 
like Re and Os compared to the 4d elements like Mo, the onsite spin anisotropy is also expected
 to be high. Let us choose our axis of quantization along this anisotropy axis, whereupon we
 need only consider the diagonal components of the last $J_{2}$ term. 
 Hence, the action can be written as:
$$\mathcal{A}=\sum_{i\omega_{n}} \left[ (i\omega_{n}-\epsilon_{Fe})\sum_{i\sigma}f_{i,n,\sigma}^{\dagger}f_{i,n,\sigma}
+t_{FM}\sum_{<ij>}(f_{i,n,\sigma}^{\dagger}m_{j,n,\sigma}+h.c.)\right.$$ 
\begin{eqnarray}
\left.+J_{1}\sum_{i\alpha\beta}\vec{S}_{i}\cdot f_{i,n,\alpha}^{\dagger}\vec{\tau}_{\alpha\beta}f_{i,n,\beta}+\sum_{i,\alpha,\beta,\delta}m_{i,n,\alpha}^{\dagger}M_{\alpha\beta}m_{i,n,\beta}\right] 
\label{actionM}
\end{eqnarray}
where the matrix $M$ is given by: $M=(i\omega_{n}-\epsilon_{Mo})\mathbf{I}+J_{2}S_{i+\delta}^{z}\sigma_{z}$.
Hence, first taking the $J_{1}\rightarrow\infty$ limit~\cite{Guinea1}, and then integrating the $B^{\prime}$ degrees of freedom out, we get the following form 
 for the action:

$$\mathcal{A}=\sum_{i,n}(i\omega_{n}-\epsilon_{Fe})\gamma_{i,n}^{\dagger}\gamma_{i,n} $$
$$-\sum_{<<i,j>>}\frac{t_{FM}^{2}}{i\omega_{n}-\epsilon_{Mo}+J_{2}\sum_{\delta}S_{\delta}^{z}} cos\frac{\theta_{i}}{2} cos\frac{\theta_{j}}{2}\gamma^{\dagger}_{i,n}\gamma_{j,n} $$ 
\begin{equation}
-\sum_{<<i,j>>}\frac{t_{FM}^{2}}{i\omega_{n}-\epsilon_{Mo}-J_{2}\sum_{\delta}S_{\delta}^{z}} sin\frac{\theta_{i}}{2} sin\frac{\theta_{j}}{2}\gamma^{\dagger}_{i,n}\gamma_{j,n}   
\label{actionJinf}
\end{equation}
where $\gamma_{i,n}$ refer to transformed spinless Fermion operators at the Iron site, while $\theta_{i}$ refer to the azimuthal angle made by the
Fe core spins with the z-axis. Following usual practice, we have neglected the Berry's phase degrees of freedom connected with the polar angles.

Hence, if we consider total energy $U=-\left<\frac{\partial A}{\partial \beta}\right>$ as before~\cite{mePinaki}, then we get:
$$ U=\sum_{<<ij>>}t_{FM}^{2}\left[\frac{(2i\omega_{n}-\Delta+J_{2}\sum_{\delta}S_{\delta}^{z})}{(i\omega_{n}-\Delta+J_{2}\sum_{\delta}S_{\delta}^{z})^{2}}cos\frac{\theta_{i}}{2}cos\frac{\theta_{j}}{2}\right.$$
\begin{equation}
+\left.\frac{(2i\omega_{n}-\Delta-J_{2}\sum_{\delta}S_{\delta}^{z})}{(i\omega_{n}-\Delta-J_{2}
\sum_{\delta}S_{\delta}^{z})^{2}}sin\frac{\theta_{i}}{2}sin\frac{\theta_{j}}{2}
\right]G_{n,ij}
\label{spinham} 
\end{equation}

So far, no approximation has been made except that $J_{2}\ll J_{1}$ and a large anistropy at
the $B^{\prime}$ site. However, to actually
calculate the exchange from this expression, one first needs to start with some spin background, and calculate the Iron Green's function in real space in this background using
 the original Hamiltonian~\ref{ham3d5d}. Then, one needs to recalculate the spin background
 from the effective spin Hamiltonian~\ref{spinham} again using some technique like Monte Carlo.
 This procedure, to be repeated till self-consistency, defines the technique of Self-consistent Renormalization~\cite{PinakiSCR}. The entire procedure is however numerically expensive and
 difficult. We shall, in the next section, instead, try to calculate the exchange analytically
 in certain simple arrangement of the background spins: namely fully ordered ferromagnetic configuration. This can act as the first step of the SCR procedure, which will be taken up in a future
work. As we shall see, it will give us an important analytic pointer at the actual magnetic
ground states possible in these compounds.

\section{Exchange in ordered spin background}

 It is to be noticed that unlike our previous SCR formulation~\cite{mePinaki} the action no longer has full spin-rotational
 invariance, owing to the anisotropy. (Notice that putting $J_{2}=0$ recovers the familiar rotationally invariant
Anderson-Hasegawa form.) Hence, even if we start with a perfectly ordered spin background with all spins parallel,  we must also choose an orientation
relative to the anisotropy axis. We consider 3 cases. In the following, we replace $\epsilon_{Fe}=0$, and $\epsilon_{Mo}=\Delta$ for convenience. 
Henceforth, $J_{2}$ will mean  $zJ_{2}$, where z is the coordination  number.

First, we consider the case $\theta_{i}=0$ for all sites.
 Then, translational invariance is restored, and we can write the relevant terms of the action \ref{actionJinf} in momentum space and calculate the
exchange as follows:
\begin{eqnarray}
{\mathcal A}=\sum_{k,n}\left(i\omega_{n}-\frac{t_{FM}^{2}}{i\omega_{n}-\Delta-J_{2}}\right)\gamma_{kn}^{\dagger}\gamma_{kn} 
\end{eqnarray}
Now we calculate $U=-\left<\frac{\partial A}{\partial \beta}\right>$,  to obtain
\begin{eqnarray}
U=\sum_{kn}\frac{t_{FM}^{2}(2i\omega_{n}-\Delta-J_{2})}{(i\omega_{n}-\Delta-J_{2})^{2}} G_{kn}
\end{eqnarray}

Evaluating the Matsubara sum,
$$J^{eff}_{ij}=\sum_{k}\frac{1}{2}\left[E_{k+}n_{F}(E_{k+)}\right.$$
\begin{eqnarray}
\left.+E_{k-}n_{F}(E_{k-})-(\Delta+J_{2})n_{F}(\Delta+J_{2})\right]e^{\vec{k}\cdot(\vec{r}_{i}-\vec{r}_{j})}
\end{eqnarray}
where $E_{k\pm}=\frac{\Delta+J_{2}\pm\sqrt{(\Delta+J_{2})^{2}+4\epsilon_{k}^{2}}}{2}$.
This gives the final expression for the exchange which has to be evaluated on a square lattice.

The case of $\theta_{i}=\pi$ gives, proceeding in a similar fashion,
$$J^{eff}_{ij}=\sum_{k,}\left[E_{k+}n_{F}(E_{k+)}\right.$$
\begin{eqnarray}
\left.+E_{-}n_{F}(E_{k-})-(\Delta+J_{2})n_{F}(\Delta+J_{2})\right]e^{\vec{k}\cdot(\vec{r}_{i}-\vec{r}_{j})}
\end{eqnarray}
The same as $\theta=0$! This is not surprising, since uniaxial anisotropy does not distinguish between $\theta=0$
 and $\theta=\pi$.
Now, let us consider the case of $\theta_{i}=\frac{\pi}{2}$.
\begin{eqnarray}
{\mathcal A}=\sum_{kn}\left[i\omega_{n}-\frac{1}{2}\frac{t_{FM}^{2}}{i\omega_{n}-\Delta}-\frac{1}{2}\frac{t_{FM}^{2}}{i\omega_{n}-\Delta}\right]\gamma_{kn}^{\dagger}\gamma_{kn}
\end{eqnarray}
Hence,
\begin{eqnarray}
U=\sum_{kn}\frac{1}{2}\left[\frac{\epsilon_{k}^{2}(2i\omega_{n}-\Delta)}{(i\omega_{n}-\Delta)^{2}}+
   \frac{\epsilon_{k}^{2}(2i\omega_{n}-\Delta)}{(i\omega_{n}-\Delta)^{2}}\right]\left<\gamma_{kn}^{\dagger}\gamma_{kn}\right>
\end{eqnarray}
Thus $J_{2}$ drops out entirely! Thus, the expression for the exchange here would coincide with that
for $J_{2}=0$ evaluated earlier~\cite{mePinaki}, which is also not surprising because the anisotropy 
axis has no component along the perpendicular direction, so that the superexchange has no effect on spin
configurations along this direction.
Replacing the Green's function,
\begin{eqnarray}
U&=&\sum_{kn}\left[\frac{(2i\omega_{n}-\Delta)}
{(i\omega_{n}-\Delta)^{2}}\right] \nonumber\\
&\times&\left(\frac{\epsilon_{k}^{2}}{i\omega_{n}-\frac{\epsilon_{k}^{2}}{(i\omega_{n}-\Delta)}}\right) \\
\end{eqnarray}
This upon evaluation of the Matsubara sum gives the familiar expression~\cite{mePinaki} 
$U=\sum_{k}\left[E_{k+}n_{F}(E_{k+})+E_{k-}n_{F}(E_{k-})-\Delta n_{F}(\Delta)\right]$ as before.
This is to be expected, since the anisotropy, and consequently, the superexchange, will have no effect in a direction perpendicular to
the easy axis. \\

Finally we write the exchange for a general $\theta_{i}=\theta$. \\
\begin{widetext}
\begin{eqnarray}
J_{ij}=t_{FM}^{2}\sum_{kn}e^{i\vec{k}\cdot(\vec{R}_{i}-\vec{R}_{j})}\left[\frac{cos^{2}\frac{\theta}{2}(2i\omega_{n}-\Delta-J_{2}cos\theta)}
{(i\omega_{n}-\Delta-J_{2}cos\theta)^{2}}+\frac{sin^{2}\frac{\theta}{2}(2i\omega_{n}-\Delta+J_{2}cos\theta)}{(i\omega_{n}-\Delta+J_{2}cos\theta)^{2}}\right]
\times G_{kn} 
\end{eqnarray}
\end{widetext}

where

\begin{eqnarray}
G_{kn}^{-1}=i\omega_{n}-\frac{\epsilon_{k}^{2}[(i\omega_{n}-\Delta)+J_{2}cos^{2}\theta]}{(i\omega_{n}-\Delta)^{2}-J_{2}^{2}cos^{2}\theta}
\end{eqnarray}

\section{Results}

\begin{figure}
\includegraphics[width=6cm]{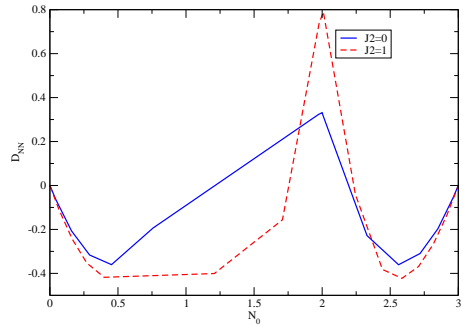}
\caption{Exchange vs filling obtained from SCR, compared between cases with and without superexchange} 
\label{fig-5}
\end{figure}

The results of the exchange calculations are shown here. For the case of $\theta=0$ (up) the exchange is as shown in
Fig~\ref{fig-5}. Since we are only interested in looking at the ferromagnetic phase, hence only the dominant nearest neighbour exchanges are shown. As explained earlier, the same graph is obtained for the $\theta=\pi$ (down) case. For comparison, the $J=0$ curve for the same $\Delta$ is also
shown~\cite{mePinaki}. This may also be thought to be the exchange data for the $\theta=\pi/2$ case. Two effects are immediately observed. Firstly, the 
extent of the ferromagnetic phase in the filling regime increases for the finite superexchange case, as compared to no superexchange. This is probably why 
for an extended filling regime, the ferromagnetic behaviour
persists rather than the antiferromagnetic behaviour, in the 3d-5d compounds. In particular, while the exchange already becomes positive for a filling
of $n\approx0.7$ for the $J_{2}=0$ case, it continues to be positive way beyond $n=1$ for finite $J_{2}$. Secondly, the magnitude of the negative part is
also larger, proving that the ferromagnetic $T_{c}$ is enhanced by superexchange. Another interesting point to note is that the exchange continues to
be large and negative till about $n=1$, which corresponds to $N=3$ in the real double perovskites owing to $t_{2g}$ degeneracy
 (eg. Sr$_{2}$CrOsO$_{6}$), while its magnitude diminishes
beyond that. The point $n=2$ shows a strong isolated antiferromagnetic tendency, possibly due to the filling of the Mo level with opposite spin in the half-metallic state.

\begin{figure}
\vspace{0.5 in}
\includegraphics[width=6cm]{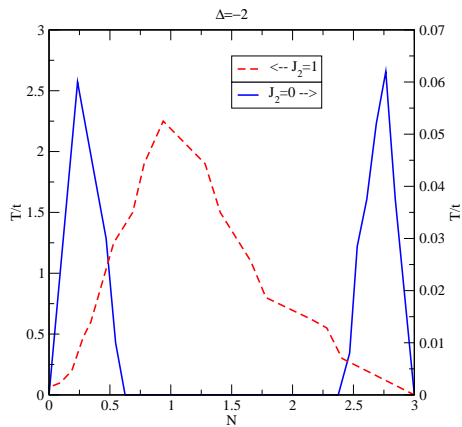}
\caption{$T_{c}$ vs N ferromagnetic part of phase diagram from 8x8 cluster compared between $J_{2}$=0 and $J_{2}$=1 ($\Delta=-2$)} 
\label{EDMC}
\end{figure}

Next, to bring these results in perspective, and establish them on a sound footing, we also performed extensive numerical simulations of the
original spin-Fermion model Hamiltonian, using the technique of exact diagonalization coupled with Monte Carlo (ED+MC). Calculations performed
in real space on a $8\times8$ cluster is shown in the figure Fig~\ref{EDMC}. We used $\Delta=-2$ and $J_{2}=1$ as before. It is found that inclusion of the $J_{2}$ term in the Hamiltonian not only results in 
a tremendous increase in the Curie temperatures, but also in the proliferation of the ferromagnetic phase to higher fillings. In fact, for these regime
of parameters, the ferromagnetic phase is found to make substantial entries into those regions of filling which were earlier reserved for the
antiferromagnetic phases~\cite{mePinaki}. However, both effects are far more augmented than suggested by the SCR calculations. Of course, the SCR is 
performed assuming ordered spin backgrounds appropriate at low temperatures and
also only one loop of the actual SCR process is performed to derive the analytical expressions given in this paper, hence it should only be considered
 as a pointer to the actual renormalization effects close to $T_{c}$. Interestingly, in the exact numerical simulations, both the ferromagnetic lobes 
at low and high filling join together in the case of finite $J_{2}$, giving a maximum $T_{c}$ at N=1 (which corresponds to N=3 considering the $t_{2g}$
degeneracy in the real material). This is consistent with the enigmatic high $T_{c}$ observed in the compound Sr$_{2}$CrOsO$_{6}$~\cite{vaithee,footnote2}. 
Also, it substantiates the view that this compound is at the threshold of a magnetic and electronic transition~\cite{vaithee}.
Thus the mechanism of superexchange coupling between B and B$^{/}$ sites provides an alternate, and more acceptable mechanism for the $T_{c}$
increase, rather than the Hubbard U mechanism invoked earlier~\cite{Guinea2}. 

\section{Summary and Outlook} 

Using the combination of effective exchange calculations within SCR and direct numerical simulations of the microscopic Hamiltonian, 
we showed that the superexchange is the main player in enhancing T$_{c}$ as well as stabilizing ferromagnetism in the Cr-compounds.
Spin-orbit coupling breaks the rotational symmetry on the B$^{'}$ site, so that different orientations of spin configurations have to be considered
 for the SCR. Amongst the configurations considered, those along the easy axis are found to increase the effective nearest neighbour ferromagnetic exchange,
 as well as contribute to the increase in the filling extent of the ferromagnetic phase. Both are signatures of stabilization of the ferromagnetic phase,
 which 
 results in the increase in $T_{c}$, and continuation of the ferro phase to higher fillings, as for Rh and Os compounds.
  Extensive numerical simulations are found to reproduce
 both of this behaviour, and provides a sound backing to the proposed model Hamiltonian as an useful one for describing the behaviour of
 Cr-based 3d-5d double perovskites.

\section{Acknowledgment}
The author gratefully acknowledges discussions with M. Randheria, D. D. Sarma, P. Majumdar and T. Saha Dasgupta.


\newpage

\begin{thebibliography}{99}
\bibitem{SFMO}K.-I. Kobayashi, T. Kimura, H. Sawada, K. Terakura, and Y. Tokura, Nature (London) {\bf 395}, 677 (1998),                            
\bibitem{DDreview} D.D. Sarma, Current Opinion in Solid State and Materials Science, {\bf 5},
261 (2001).
\bibitem{tomioka} Y. Tomioka, T. Okuda, Y. Okimoto, R. Kumai, K.-I. Kobayashi, Y. Tokura, Phys. Rev. B {\bf 61}, 422 (2000).
\bibitem{TSD}D.D. Sarma, P. Mahadevan, T. Saha Dasgupta, S. Ray, A. Kumar, Phys. Rev. Lett., {\bf 85}, 2549 (2000).
\bibitem{Mag-Res1}B. Garcia Landa et al, Solid State Comm., {\bf 110}, 435 (1999).
\bibitem{Mag-Res2} B. Martinez, J. Navarro, L. Balcells and J. Fontcuberta, J. Phys.: Condens. Matter {\bf 12}, 10515 (2000).
\bibitem{Mag-Res3}D. D. Sarma, S. Ray, K. Tanaka, M. Kobayashi, A. Fujimori, P. Sanyal, H.R. Krishnamurthy and C. Dasgupta,
Phys. Rev. Lett., {\bf 98}, 157205 (2007).
\bibitem{JAP}A. Kahoul, A. Azizi, S. Colis, D. Stoeffler, R. Moubah, G. Schmerber, C. Leuvrey, and A. Dinia, J. Appl. Phys. {\bf 104}, 123903 (2008).
\bibitem{ES}T. Saitoh, M. Nakatake, H. Nakajima, O. Morimoto, A. Kakizaki, Sh. Xu, Y. Moritomo, N. Hamada, Y. Aiura, Journal of Electron Spectroscopy and Related Phenomena {\bf 144-147}, 601 (2005).
\bibitem{Navarro} J. Navarro, C. Frontera, Ll. Balcells, B. Mart$\acute{i}$nez, and J. Fontcuberta, Phys. Rev. B 
{\bf 64}, 092411 (2001).
\bibitem{p21n} Carlos Frontera, Diego Rubi, Jose Navarro, Jose Luis Garcia-Munoz, and Josep Fontcuberta, Phys. Rev. B {\bf 68}, 012412 (2003).
\bibitem{photoemission} J. Navarro, J. Fontcuberta, M. Izquierdo, J. Avila, M.C. Asensio, Phys. Rev. B,{\bf 70},054423 (2004).
\bibitem{mePinaki}Prabuddha Sanyal and Pinaki Majumdar, Phys. Rev. B, {\bf 80}, 054411 (2009).
\bibitem{LSFMO} P. Sanyal, H.Das and T. Saha Dasgupta, Phys. Rev. B, {\bf 80}, 224412 (2009).
\bibitem{Sugatapriv} S. Jana, C. Meneghini, P. Sanyal, S. Sarkar, T. Saha-Dasgupta, S. Ray, submitted to Phys. Rev. Lett. (2012). 
\bibitem{3d5d} Hena Das, Prabuddha Sanyal, T. Saha-Dasgupta and D.D. Sarma, Phys. Rev. B, {\bf 83}, 104418 (2011)
\bibitem{GuyCohen} G. Cohen, V. Fleurov, K.Kikoin, J. Appl.Phys., {\bf 101}, 09H106 (2007).
\bibitem{NMTO}O. K. Andersen and T. Saha-Dasgupta, Phys. Rev. B {\bf 62}, R16219 (2000).

\bibitem{VASP}G. Kresse and J. Hafner, Phys. Rev. B {\bf 47}, R558 (1993), G. Kresse and J. Furthmueller, Phys. Rev. B {\bf 54}, 11169 (1996).
\bibitem{Millis}A. Chattopadhyay and A. J. Millis, Phys. Rev. B {\bf 64}, 024424 (2001).
\bibitem{Avignon}O. Navarro, E. Carvajal, B. Aguilar, M. Avignon, Physica B {\bf 384}, 110 (2006).
\bibitem{Guinea2}L. Brey, M. J. Calder$\acute{o}$n, S. Das Sarma and F. Guinea, Phys. Rev. B {\bf 74}, 094429 (2006).
\bibitem{Guinea1} J.L.Alonso,L.A. Fernandez, F. Guinea, F. Lesmes, and V. Martin-Mayor, Phys. Rev. B, {\bf 67}, 
   214423 (2003).
 
\bibitem{PinakiSCR} S. Kumar and P. Majumdar, Eur. Phys. J. B, {\bf 46}, 315 (2005). 

\bibitem{GGA}J. P. Perdew, J. A. Chevary, S. H. Vosko, K. A. Jackson, M. R. Pederson, D. J. Singh, and C. Fiolhais, Phys. Rev. B {\bf 46}, 6671 (1992); {\bf 48}, 4978(E) (1993).
\bibitem{vaithee} Y. Krockenberger, K. Mogare, M. Reehuis, M. Tovar, M. Jansen, G. Vaitheeswaran, V. Kanchana,
                  F. Bultmark, A. Delin, F. Wilhelm, A. Rogalev, A. Winkler and L. Alff, Phys. Rev. B, {\bf 75},
   020404 (2007).
\bibitem{footnote} In real double perovskites, due to the threefold $t_{2g}$ degeneracy, there are a total of nine degrees of freedom per Fe-Mo pair, consisting of three at Fe site and six at Mo site. Hence, to compare
with experiments, our x-axes should be multiplied by 3.
\bibitem{footnote2} We have deliberately considered a larger value of $J_{2}$ than usually found in real Cr-based DP-s to underline the strong
  $T_{c}$ enhancement and ferromagnet-proliferation effects. Actual values of $J_{2}$ may leave an window for the
  antiferromagnetic phases as well. The effects of phase competition in this Hamiltonian will be considered in a later work.
 


\end{thebibliography}
\end{document}